\documentclass[aps,pre,twocolumn,groupedaddress,showpacs,floatfix]{revtex4}
\usepackage{graphicx}
\usepackage{pst-all}
\usepackage{color}
\usepackage{amsmath}
\newcommand{\comment}[1]{}

\begin{document}
\title{On the number of attractors in random Boolean networks}
\author{Barbara Drossel}
\affiliation{Institut f\"ur Festk\"orperphysik,  TU Darmstadt,
Hochschulstra\ss e 6, 64289 Darmstadt, Germany }
\date{\today}
\begin{abstract}
The evaluation of the number of attractors in Kauffman networks by
Samuelsson and Troein is generalized to critical networks with one
input per node and to networks with two inputs per node and different
probability distributions for update functions. A connection is made
between the terms occurring in the calculation and between the more
graphic concepts of frozen, nonfrozen and relevant nodes, and relevant
components. Based on this understanding, a phenomenological argument
is given that reproduces the dependence of the attractor numbers on
system size.
\end{abstract}
\maketitle

\section{Introduction}
\label{intro}

Boolean networks are often used as generic models for the dynamics of
complex systems of interacting entities, such as social and economic
networks, neural networks, and gene or protein interaction networks
\cite{kauffman:random}. The simplest and most widely studied of these
models was introduced in 1969 by Kauffman \cite{kauffman:metabolic} as
a model for gene regulation.  The system consists of $N$ nodes, each
of which receives input from $K$ randomly chosen other nodes. The
network is updated synchronously, the state of a node at time step $t$
being a Boolean function of the states of the $K$ input nodes at the
previous time step, $t-1$.  The Boolean updating functions are
randomly assigned to every node in the network, and together with the
connectivity pattern they define the realization of the network. For
any initial condition, the network eventually settles on a periodic
attractor. Thus the number and the lengths of the attractors are
important features of the networks. Of special interest are
\emph{critical} networks, which lie at the boundary between a frozen
phase and a chaotic phase \cite{derrida:random,derrida:phase}.  In the
frozen phase, a perturbation at one node propagates during one time
step on an average to less than one node, and the attractor lengths
remain finite in the limit $N\to \infty$. In the chaotic phase, the
difference between two almost identical states increases exponentially
fast, because a perturbation propagates on an average to more than one
node during one time step \cite{aldana-gonzalez:boolean}. Based on
computer simulations, the mean attractor number of critical $K=2$
Kauffman networks with a constant probability distribution for the 16
possible updating functions was once believed to scale as $\sqrt{N}$
\cite{kauffman:metabolic}. With increasing computer power, a faster
increase was seen (linear in \cite{bilke:stability}, ``faster than
linear'' in \cite{socolar:scaling}, stretched exponential in
\cite{bastolla:relevant,bastolla:modular}). Then, in a beautiful
analytical study, Samuelsson and Troein
\cite{samuelsson:superpolynomial} have proven that the number of
attractors grows indeed faster than any power law with the network
size $N$. A proof that the number and length of attractors of critical
$K=1$ networks increases faster than any power law was published some
time later \cite{drossel:number}. These two proofs, although they
apply to closely related systems, are conceptually
different. The latter derives structural properties of the relevant
part of the networks, and obtains from there a lower bound for the
number of attractors. In contrast, in
\cite{samuelsson:superpolynomial} the mean number of attractors is
obtained by a direct calculation that uses the saddle-point
approximation, and which yields no graphic understanding of how the
attractor numbers arise. 

It is the purpose of the present article to show how the attractor
numbers arise in terms of the relevant parts of the networks. To this
aim, the method chosen by Samuelsson and Troein is in the next section
applied to the critical $K=1$ networks, for which an intuitive
understanding already exists. The dependence of attractor numbers on
system size $N$ can indeed be reproduced by phenomenological arguments
based on this understanding.  In section \ref{general}, it is shown
that these networks are similar in many respects to critical $K=2$
networks, of which a more general class than usual will be
defined. Applying the calculation to this more general class leads
eventually to a phenomenological argument that reproduces the
dependence of attractor numbers on system size.

\section{Critical networks with one input per node}
\label{k1}

Let us first consider critical networks with connectivity $K=1$. A
random network with one input per node is critical if among the four
possible Boolean functions only the two nonfrozen ones, ``copy'' and
``invert'', are chosen.  In \cite{flyvbjerg:exact} and
\cite{drossel:number}, exact results for the topology of $k=1$
networks are derived. The network consists of the order of $\ln (N)$
unconnected components, each of which contains a loop of
\emph{relevant nodes}, and trees rooted in these loops.  Relevant
nodes are defined as those nodes whose state is not constant and that
control at least one relevant element \cite{bastolla:modular}. They
determine the attractors of the system.  The number of loops of size
$l$ is Poisson distributed with a mean $1/l$, if $l$ is smaller than a
cutoff size $l_c$. The cutoff loop size scales as $\l_c \sim \sqrt{N}$
\cite{flyvbjerg:exact,drossel:number}.

Following the calculation by Samuelsson and Troein
\cite{samuelsson:superpolynomial}, we calculate in the following the
mean number of attractors of length $L$. More precisely, we calculate
the mean number of cycles in state space. While an attractor is always
a cycle in state space, the reverse is not necessarily true, since an
attractor must be accompanied by a shrinking state space volume.
However, for the networks discussed in this paper, cycles are almost
always attractors, since the dynamics on the trees rooted in the loops
is being slaved to the dynamics on the loops, and therefore the
initial states of the trees will be forgotten. For every network that
contains trees, the number of initial states that leads to a given
cycle is larger than the period of the cycle, and the cycles are
attractors.

Let $\langle C_L\rangle_N$ denote the mean number of cycles in state space
of length $L$, averaged over the ensemble of networks of size $N$. On
a cycle of length $L$, the state of each node goes through a sequence
of 1s and 0s of period $L$. Let us number the $2^{L-1}$ possible
sequences of period $L$ of the state of a node by the index $j$,
ranging from 0 to $m-1\equiv 2^{L-1}-1$, with sequence 0 being the
constant one. Following Samuelsson and Troein, we consider two
sequences as identical if they can be transformed into each other by
exchanging 1s and 0s. This simplifies the calculation a lot, since the
sequence of the node from which a node with sequence $j$ receives its
input, can only be one sequence, which we denote $\phi(j)$. It is
obtained from $j$ by taking the first bit of $j$ and moving it to the
end of the sequence. Whether the Boolean function at a node is
``copy'' or ``invert'', has now become irrelevant, and all results
obtained in this section apply therefore to critical $K=1$ networks
with a proportion $p$ of ``copy''  functions and a proportion $1-p$ of
``invert'' functions, for any value of $p$. 

If $n_j$ is the number of nodes that have the sequence $j$ on a cycle
of length $L$, and $\mathbf{n}$ the vector $(n_0,\ldots,n_{m-1})$, then
\begin{equation}
\langle C_L\rangle_N = \frac 1 L \sum_{\mathbf{n}} {N \choose
\mathbf{n}} \prod_{j=0}^{m-1}\left(\frac {n_{\phi(j)}}N\right)^{n_j}\, ,
\label{my3}
\end{equation}
where $N \choose \mathbf{n}$ denotes the multinomial
$N!/(n_0!\dots n_{m-1}!)$, i.e., the number of different ways to assign
the sequences 0 to ${m-1}$ to $n_0,\dots, n_{m-1}$ nodes. The factor $1/L$
occurs because any of the $L$ states on the cycle could be the
starting point, and the product is the probability that each node with
a sequence $j$ is connected to  a node with the sequence
$\phi(j)$. For sufficiently large $N$, all $n_j$ will be large, and we
can apply Stirling's formula $n_j! = (n_j/e)^{n_j}\sqrt{2\pi
  n_j}$. Transforming the variables from $\mathbf{n}$ to $\mathbf{x} =
\mathbf{n}/N$, we can replace the sum with an integral and obtain
\begin{equation}
\langle C_L\rangle_N \simeq \frac 1 L \left(\frac N
	{2\pi}\right)^{\frac{m-1} 2}
\int d\mathbf{x} \frac{e^{N\sum_j x_j \ln(x_{\phi(j)}/x_j)}}{\prod_{j=0}^{m-1}
	\sqrt{x_j}}\, .\label{my6}
\end{equation}
Itegration space is limited by the condition $\sum_j x_j =1$ (with all
$x_j>0$). The integral is evaluated using the saddle-point
approximation, which becomes exact in the thermodynamic limit $N \to
\infty$. The maximum of the expression $ \sum_j x_j
\ln(x_{\phi(j)}/x_j)$ is obtained when $x_{\phi(j)} = x_j$ for all
$j$. This means that the space of sequences $j$ is decomposed into
permutation sets of the type $\{j,\phi(j), \phi(\phi(j)),\dots\}$,
with all members of a set occurring equally often at the saddle
point. This can be understood from the topological structure of $K=1$
networks. All nodes that are on the same component, must undergo a
sequence belonging to the same set, while different components are
independent from each other. Furthermore, on a loop or an infinitely
long line of nodes, every member of the set occurs equally often,
since between nodes with identical sequences, there must be nodes with
all the other sequences from the set. The deviation from $x_{\phi(j)}
= x_j$ evaluated below comes from the fact that the branches of the
trees have a finite length, which is generally not a multiple of the
set size. 

Let the index $h$ count the permutation sets, with $h=0,\dots,H_L-1$.
Let $\rho_L^h$ be the set with index $h$, which has $|\rho_L^h|$
members. In order to perform the saddle point integration, we make a
transformation of variables within each set, defining $z_h=\sum_{j\in
\rho_L^h} x_j$, and $\delta x_j = x_j - z_h/|\rho_L^h|$, with
$\sum_{j\in \rho_L^h}\delta x_j = 0$. Only $|\rho_L^h|-1$ of all the
$\delta x_j$ within a set are independent.

Expanding to second order in the $\delta x_j$, we obtain for the exponent in (\ref{my6})
\begin{eqnarray}
\sum_{j\in \rho_L^h}  x_j \ln\frac{x_{\phi(j)}}{x_j} &=& \sum_{j\in
  \rho_L^h}
\left(\frac{z_h}{|\rho_L^h|} + \delta x_j\right) \ln
  \left(\frac{1+\frac{|\rho_L^h|}{z_h}\delta x_{\phi(j)}
  }{1+\frac{|\rho_L^h|}{z_h}\delta x_{j}}\right)\nonumber\\
&\simeq& 
% \sum_{i\in \rho_L^h} \left(\frac{z_h}{|\rho_L^h|} + \delta x_i\right)
%  \left[\frac{|\rho_L^h|}{z_h}\delta x_{\phi(j)} -
%  \left(\frac{|\rho_L^h|}{z_h}\delta x_{\phi(j)}\right)^2 -
%  \frac{|\rho_L^h|}{z_h}\delta x_i -
%  \left(\frac{|\rho_L^h|}{z_h}\delta x_i\right)^2\right]\nonumber\\
%&=&  
\frac{|\rho_L^h|}{z_h} \sum_{j\in
  \rho_L^h}\delta x_j (\delta x_{\phi(j)} - \delta x_j) \nonumber\\
&=& -\frac 1 2  \frac{|\rho_L^h|}{z_h} \sum_{j\in
  \rho_L^h} (\delta x_{\phi(j)} - \delta x_j)^2
\end{eqnarray}
and
$$ \prod_{i\in \rho_L^h} (x_j)^{-1/2} \simeq
\left(\frac{z_h}{|\rho_L^h|}\right)^{-|\rho_L^h|/2}\, .$$
In the last equation, terms containing $(\delta x_j)^2$ vanish in the limit
$N\to \infty$, since the saddle-point integration gives contributions
only from values $\delta x_j$ of the order of $1/\sqrt N$. 

The integral over the $\delta x_j$ can be performed by using the
variables $ (\delta x_{\phi(j)} - \delta x_j)$, leading to
\begin{eqnarray}
\langle C_L\rangle_N& \simeq& \frac 1 L \left(\frac N
	{2\pi}\right)^{\frac{m-1}2} \prod_h \left[
\int \frac{
 \frac{dz_h}{|\rho_L^h|}
}{
\left(\frac{z_h}{|\rho_L^h|}\right)^{\frac{|\rho_L^h|}{2}}
}
\left(\frac{2\pi z_h}{{|\rho_L^h|}N}\right)^{\frac{|\rho_L^h|-1}{2}}\right] \nonumber\\
&=&  \frac 1 L \left(\frac N {2\pi}\right)^{\frac{H_L-1}2}\prod_h \left[\frac 1
     {\sqrt{|\rho_L^h|}}
\int dz_h \frac 1 {\sqrt{z_h}}\right]\, .\label{myresult}
\end{eqnarray} 
Integration space is given by the condition $\sum_h z_h =1$ (with all
$z_h>0$). 

Let us now interpret the $N$-dependence in this result. To this
purpose, we derive the number of attractors of length $L$ from the
known topological properties of $K=1$ networks. As mentioned above,
the network consists of the order of $\ln N$ components, each of which
contains a loop and trees rooted in the loops. The cutoff in loop size
is $l_c \sim \sqrt{N}$. The expected number of states on a loop of a randomly chosen size
$l$ that belong to a cycle of length $L$ is
\begin{equation}
\sum_h \frac 1{|\rho_L^h|} |\rho_L^h| = H_L\, , \label{needed}
\end{equation}
 the first factor
being the probability that $l$ is a multiple of $|\rho_L^h|$, and the
second factor being the number of states in the set number $h$. As
mentioned above, the number $n_l$ of loops of size $l$ is Poisson
distributed with a mean $1/l$, leading to
\begin{eqnarray}
\langle C_L\rangle_N &\simeq &
\sum_{\{n_l\}}\prod_{l \le l_c} \left(\frac{e^{-1/l} \left(\frac 1
  {l}\right)^{n_l}}{n_l!} H_L^{n_l}\right)\nonumber\\
&=& \sum_{\{n_l\}}\prod_{l \le l_c} \left(\frac{e^{-1/l} \left(\frac {H_L}
  {l}\right)^{n_l}}{n_l!} \right)\nonumber\\
&\simeq& \prod_{l \le l_c} e^{(H_L-1)/l} = e^{(H_L-1)\int_1^{l_c} dl/l}
  \nonumber\\
&\sim&  e^{(H_L-1)\ln\sqrt N} = N^{(H_L-1)/2} \, .\label{quick}
\end{eqnarray}
The mean number of attractors of length $L$ scales as the number of
relevant nodes, $\sqrt{N}$ (which is proportional to the number of
nodes in the largest loop), to the power $H_L-1$. We will see below
that an equivalent statement can be made for the $K=2$ critical
networks.  In order to obtain also the $L$-dependent prefactor in
Eq.~(\ref{myresult}), the full probability distribution of the number
of loops of a size of the order of $l_c$ would have to be taken into
account in calculation (\ref{quick}), instead of simply integrating
up to $l_c$. 

Let us conclude this section by discussing the implications of the
fact that we do not discriminate between sequences that can be
transformed into each other by exchanging 1s and 0s. The numbers and
the periods of the attractors are determined by the loops in the
network. We call a loop ``even'' if it contains an even number of
``invert'' functions, and ``odd'' if it contains an odd number of
``invert'' functions. The state of an even loop of size $l$ is the
same after $l$ updates, while the state of an odd loop of size $l$ is
inverted after $l$ updates. If $l$ is a prime number, the period of a
cycle on an odd loop is $2l$, with the second half of the cycle being
obtained from the first half by exchanging 0s and 1s. However, our
rules defined above (and in \cite{samuelsson:superpolynomial})
classify this as a cycle of period $l$, since assigning only the first
half of a sequence to the nodes on the loop, makes a contribution to
Eq.~(\ref{my3}) if $l$ is a multiple of $L$.  Furthermore, exchanging
the 1s and 0s on a component does not lead to a new cycle according to
our calculation, but in reality this doubles the number of
cycles. 

Repeating calculation (\ref{quick}) for a system with only ``copy''
functions and by discriminating sequences that can be transformed into
each other by exchanging 1s and 0s, the result remains the same, but
with $H_L$ now counting the true number of invariant sets.

For a system that contains also ``invert'' functions, the calculation
becomes more complicated, since the mean number of states of a loop
belonging to a cycle of length $L$ is no longer given by
Eq.~(\ref{needed}). Let $H_L$ again count the number of true invariant
sets. The probability that a loop has a given cycle is now
$1/2|\rho_h^L|$ if the second half of the cycle is not obtained from
the first half by exchanging 1s and 0s. Otherwise, the probability is
$3/2|\rho_h^L|$.  The mean number of states on a loop that belong to a
cycle of length $L$ is therefore $H_L/2$ for odd $L$ and $H_L/2 +
H_{L/2}$ for even $L$, and these two expressions replace the $H_L$ in
the exponent in (\ref{quick}) for odd and even $L$ respectively.

\section{A general class of critical $K=2$ networks}
\label{general}

Now, let us consider $K=2$ networks, where each node has 2 randomly
chosen inputs. The 16 possible update functions are shown in table
\ref{tab1}.

\begin{table}
\begin{center}
\begin{tabular}{|c||c|c||c|c|c|c||c|c|c|c|c|c|c|c||c|c|}\hline
In&
\multicolumn{2}{|c||}{$\mathcal{F}$}&
\multicolumn{4}{|c||}{${\mathcal{C}}_1$}&
\multicolumn{8}{|c||}{${\mathcal{C}}_2$}&
\multicolumn{2}{|c|}{$\mathcal{R}$}\\\hline
00&1&0&0&1&0&1&1&0&0&0&0&1&1&1&1&0\\
01&1&0&0&1&1&0&0&1&0&0&1&0&1&1&0&1\\
10&1&0&1&0&0&1&0&0&1&0&1&1&0&1&0&1\\
11&1&0&1&0&1&0&0&0&0&1&1&1&1&0&1&0\\\hline
\end{tabular}
\end{center}
\caption{The 16 update functions for nodes with 2 inputs. The first
  column lists the 4 possible states of the two inputs, the other
  columns represent one update function each, falling into four classes.}
\label{tab1}
\end{table}
The update functions fall into four classes
\cite{aldana-gonzalez:boolean}. In the first class, denoted by
$\mathcal{F}$, are the frozen functions, where the output is fixed
irrespectively of the input. The class ${\mathcal{C}}_1$ contains
those functions that depend only on one of the two inputs, but not on
the other one. The class ${\mathcal{C}}_2$ contains the remaining
canalyzing functions, where one state of each input fixes the
output. The class $\mathcal{R}$ contains the two reversible update
functions, where the output is changed whenever one of the inputs is
changed. Critical networks are those where a change in one node
propagates to one other node on an average. A change propagates with
probability $1/2$ to a node that has a canalyzing update function
$\mathcal{C}_1$ or $\mathcal{C}_2$, with probability zero to a node
that has a frozen update function, and with probability 1 to a node
that has a reversible update function. Consequently, if the frozen and
reversible update functions are chosen with equal probability, the
network is critical. Usually, only those models are considered where
all 16 update functions receive equal weight. We now consider the
larger set of models where the frozen and reversible update functions
are chosen with equal probability, and where the remaining probability
is divided between the $\mathcal{C}_1$ and $\mathcal{C}_2$
functions. Those networks that contain only $\mathcal{C}_1$ functions
are different from the remaining ones.  Since all nodes respond only
to one input, the link to the second input can be cut, and we are left
with a critical $K=1$ network, which was discussed in the previous
section. We shall see below that all the other models, where the
weight of the $\mathcal{C}_1$ functions is smaller than 1, fall into
the same class, where the number of attractors is given by the
expression derived in \cite{samuelsson:superpolynomial} and reproduced
below.

For all these critical $K=2$ networks, the mean number of attractors
of length $L$ is given by the expression
\cite{samuelsson:superpolynomial}
\begin{equation}
\langle C_L\rangle_N = \frac 1 L \sum_{\mathbf{n}} {N \choose \mathbf{n}} \prod_j\left(\sum_{l_1,k} \frac{n_{l_1}n_{k}}{N^2}(P_L)_{l_1,k}^j\right)^{n_j}\, ,
\end{equation}
with $(P_L)_{l_1,k}^j$ being the probability that a node that has
the input sequences $l_1$ and $k$ generates the output sequence
$j$. This expression is the obvious generalization of Eq.~(\ref{my3})
to two inputs per node. Using again Stirling's formula and replacing
the sum with an integral, this leads to the generalization of
Eq.~(\ref{my6}), see  \cite{samuelsson:superpolynomial}
\begin{equation}
\langle C_L\rangle_N \simeq \frac 1 L \left(\frac N
	{2\pi}\right)^{\frac{m-1}2}
\int d\mathbf{x} \frac{e^{N\sum_i x_i \ln\left(\frac 1 {x_i}
	\sum_{j,k} x_{j}x_{k} (P_L)^i_{jk}\right)}}{\prod_i
	\sqrt{x_i}}\, .\label{st6}
\end{equation}
For a network with only $\mathcal{C}_1$ functions, this reduces immediately
 to  Eq.~(\ref{my3}). The exponent has its maximum at zero, and this
 value is reached only if  \cite{samuelsson:superpolynomial}
\begin{equation}x_i = \sum_{j,k} x_{j}x_{k} (P_L)^i_{jk} \hbox{ for all } i\, .
\label{condition}
\end{equation}
This condition is satisfied for $x_0=1$. For a network with only
$\mathcal{C}_1$ functions, it is more generally satisfied for 
$x_{\phi(i)} = x_i$ (for all $i$). For all other critical networks,
  there exists only the maximum at $x_0=1$. This is shown as follows:
Eq.~(\ref{condition})
can be transformed into
\begin{eqnarray}
\sum_{i>0} x_i &=& \sum_{i>0}x_0^2 (P_L)^i_{00} + 2\sum_{i,j>0} x_{0}x_j
(P_L)^i_{j0}\nonumber\\
&&+  \sum_{i,k,j>0} x_{j}x_{k}
(P_L)^i_{jk}\, ,\nonumber\\
1-x_0 &=& 2x_0\sum_{j>0}\frac 1 2 x_{j} + \sum_{j,k,j>0}
x_{j}x_{k}(P_L)^j_{jk}\nonumber\\
&\le& x_0(1-x_0) + (1-x_0)^2 = 1-x_0\, .
\end{eqnarray}
Here, we have used $(P_L)^j_{00} = 0$ for $j>0$ and
$\sum_{j>0}(P_L)^j_{i0}=1/2$ for all considered models.  The
inequality becomes an equality only if $x_0=1$, or if
$\sum_{i,k,j>0}x_{j}x_{k}(P_L)^i_{jk} = (1-x_0)^2$. The
latter condition is satisfied if and only if all $(P_L)^0_{jk}$
with $j,k>0$ and $x_{j},x_{k}>0$ vanish. They cannot vanish if
there are frozen update functions. They do vanish if there are only
$\mathcal{C}_1$ update functions. It remains to be shown that they
cannot vanish for a system containing $\mathcal{C}_2$ functions, but
no frozen functions. Assume $x_i>0$. Since a node with two input
sequences $i$ has nonconstant output sequences (one of which we denote
by $ii$) with a positive probability, there occurs a term
$x_ix_{ii}(P_L)^j_{i,ii}$. Now, the sequences $i$ and $ii$ taken
together, have only 2 out of the 4 possible combinations of 2
bits. However, among the $\mathcal{C}_2$ functions there are
functions that yield a constant output if the input is $i$ and
$ii$. Therefore even in a $\mathcal{C}_2$ network, not all
$(P_L)^0_{jk}$ with $j,k>0$ and $x_{j},x_{k}>0$ vanish. We
  thus have shown that all considered critical $K=2$ networks 
satisfy (\ref{condition}) only at $x_0=1$. For large $N$, only small values
$x_j$ (for $j>0$) contribute to the integral in (\ref{st6}), and a
Taylor expansion in the $x_j$ (for $j>0$) gives
\cite{samuelsson:superpolynomial}
\begin{equation}
\langle C_L\rangle_N \simeq \frac 1 L \left(\frac N
	{2\pi}\right)^{\frac m 2}
\int d\mathbf{x} e^{Nf(\mathbf{x})}\end{equation}
with
\begin{eqnarray}
f(\mathbf{x}) &\simeq& \sum_{i>0} x_i \ln\frac{x_{\phi(i)}}{x_i} + \sum_{i} x_i
\frac{\mathbf{x} \cdot A_L^i \mathbf{x}}{x_{\phi(i)}} \nonumber\\
&& - \frac 1 2 \sum_{i>0} x_i \left(\mathbf{x} \cdot A_L^i
  \mathbf{x}\right)^2 \label{st9}
\end{eqnarray}
where $(A_L^i)_{jk} = (P_L)^j_{jk}-\frac 1 2
(\delta_{j\phi(i)}+\delta_{k\phi(i)})$. For a $\mathcal{C}_1$
network, the matrix $(A_L^i)$ vanishes, and we obtain again
(\ref{my6}).  
The maximum of $f(\mathbf{x})$ is obtained when $x_{\phi(i)} = x_i$
for all $i$. At this maximum, the first and second term vanish, and
the third term is of the form $N x^3$. Consequently, only values $x_i$
(with $i>0$) up to the order $N^{-1/3}$ contribute to $\langle
C_L\rangle_N $. This means that the proportion of nodes that are not
frozen on an attractor is of the order $N^{-1/3}$, and the total
number of nonfrozen nodes is of the order $N^{2/3}$. This is in
contrast to the critical $K=1$ network, where a nonvanishing
proportion of nodes is nonfrozen. Changing the variables again to 
$z_h=\sum_{i\in
  \rho_L^h} x_i$, and $\delta x_i = x_i - z_h/|\rho_L^h|$, the
integration over the  $\delta x_i$ gives now
\begin{equation}
\langle C_L\rangle_N \simeq \frac 1 L  \left(\frac N {2\pi}\right)^{\frac{H_L-1}{2}}\prod_{h>0}\left[ \frac {\int dz_h} 
     {\sqrt{|\rho_L^h|z_h}}\right]
 e^{- \sum_{h>0}\frac {N\left(\mathbf{z} \cdot B_L^h
  \mathbf{z}\right)^2 } {2z_h} }\label{mysam12}
\end{equation}
with $(B_L^h)_{jk} = (P_L')^h_{jk}-\frac 1 2 (\delta_{j h}+
\delta_{k h})$, and with $(P_L')^h_{jk}$ being the probabity
that the output sequence belongs to set $h$ if the input sequences
belong to the sets $j$ and $k$. 
Introducing a new variable $y_h = z_h N^{1/3}$, we obtain an
additional factor $N^{-(H_L-1)/6}$, and the mean number of cycles of
length $L$ becomes \cite{samuelsson:superpolynomial}
\begin{equation}
\langle C_L\rangle_N \simeq \frac 1 L  \frac {N^{\frac{H_L-1}{3}}}{ (2\pi)^{\frac{H_L-1}{2}}}\prod_{h>0}\left[ \frac {\int dy_h} 
     {\sqrt{|\rho_L^h|y_h}}\right]
 e^{- \sum_{h>0}\frac  {\left(\mathbf{y} \cdot B_L^h
  \mathbf{y}\right)^2 }{2y_h} }\label{sam12}
\end{equation}
While integration space for the $z_h$ was restricted by the condition
$\sum_h z_h=1-x_0$, there is no constraint for the $y_h$.

With the understanding gained from the $K=1$ critical networks, we can
interpret the calculation as follows. The difference between $K=1$ and
$K=2$ critical networks comes from the fact that in the $K=2$ networks
only the fraction $N^{-1/3}$ of nodes is nonfrozen. This modifies the
exponent of $N$ in the final result, and this leads to the different form of
the $z_h$ integration. Both types of networks have in common that the main
contribution to the integral comes from the neighborhood of the
subspace satifsying $x_{\phi(i)} = x_i$ for all $i$. This means that
the majority of nonfrozen nodes receive input from one nonfrozen node,
the other input being frozen. The nonfrozen part of a $K=2$ critical
network resembles therefore a $K=1$ critical network. The
proportion of nonfrozen nodes receiving input from two nonfrozen
nodes, cannot be larger than of the order $N^{-1/3}$, since the
$\delta x_i$ are of the order $N^{-1/3}$. Thus, the nonfrozen part of
a $K=2$ critical network differs from a $K=1$ critical network by a 
proportion $N^{-1/3}$ of nonfrozen nodes having two nonfrozen
inputs. Apparently, this difference does not affect the scaling of
$\langle C_L\rangle_N $ with $N$, but only the $L$-dependent
prefactor. If the number of relevant nodes scales as $N^{1/3}$ (as is
numerically found in \cite{socolar:scaling}), the law
$$\langle C_L\rangle_N \sim N^{(H_L-1)/3}$$ means that the mean number
of attractors of length $L$ scales as the number of relevant nodes,
$N^{1/3}$ (which is proportional to the number of nodes in the largest
component), to the power $H_L-1$. It can be obtained by a
phenomenological argument similar to the one used in the previous
section. Assume there are $N^{1/3}$ relevant nodes arranged in $\sim
\ln N$ components, with the number of components of size $l$ being
Poisson distributed with a mean $1/l$.  Since at most the proportion
$N^{-1/3}$ of nonfrozen nodes have two nonfrozen inputs, only a finite
number of relevant nodes have two nonfrozen inputs, and all relevant
components apart from a finite number are cycles without additional
links, just as for the $K=1$ critical network. 
 The mean number of states on a component of size
$l$ that belong to a cycle of length $L$ is therefore in the limit of
 large $N$
$$
\sum_h \frac 1{|\rho_L^h|} |\rho_L^h| = H_L\, , 
$$ just as for the $K=1$ critical network.  If the number $n_l$ of
relevant components of size $l<l_c\sim N^{1/3}$ is Poisson distributed
with a mean $1/l$, we obtain
\begin{eqnarray}
\langle C_L\rangle_N &\simeq &
\sum_{\{n_l\}}\prod_{l \le l_c} \left(\frac{e^{-1/l} \left(\frac 1
  {l}\right)^{n_l}}{n_l!} H_L^{n_l}\right)\nonumber\\
&\sim&  e^{(H_L-1)\ln N^{1/3}} = N^{(H_L-1)/3} \, .\label{quick2}
\end{eqnarray}

\section{Conclusions}
\label{conclusions}

In this paper, we have considered the mean number of attractors of
length $L$ for critical $K=1$ and $K=2$ networks. We have applied the
method by Samuelsson and Troein \cite{samuelsson:superpolynomial} and
have interpreted the results in terms of the topological properties of
the nonfrozen part of the network. For the $K=1$ networks, the
dependence of the number of attractors of length $L$ on the system
size $N$, $\langle C_L\rangle_N \sim N^{(H_L-1)/2}$ could be
understood as resulting from the network containing of the order
of $N^{1/2}$ relevant nodes arranged in $\sim \ln N$ components, with
the number of components of size $l$ being Poisson distributed with a
mean $1/l$. The nonrelevant nodes sit in trees rooted in the loops. 

Then, we could show that all $K=2$ critical networks can be treated by
the same calculation. Only for networks consisting only of
$\mathcal{C}_1$ functions, the step from Eq.~(\ref{mysam12}) to
Eq.~(\ref{sam12}) cannot be made, since the matrix $(B_L^h)_{jk}$
vanishes in this case. $\mathcal{C}_1$-networks are in fact $K=1$
critical networks, and Eq.~(\ref{mysam12}) is identical to
Eq.~(\ref{myresult}) in this case. All the other $K=2$ networks show
the same dependence of attractor numbers on system size, with only the
$L$-dependent prefactor being different (because the matrix
$(B_L^h)_{jk}$ is different for a different choice of weights for the
update functions).  We saw that only the proportion $N^{-1/3}$ of
nodes is nonfrozen, and that almost all nonfrozen nodes depend only on
one nonfrozen input. The nonfrozen part of critical $K=2$ networks
resembles strongly a $K=1$ critical network, and by analogy to the
$K=1$ critical network we concluded that 
 the scaling with $N$ of the number of
attractors of length $L$ in $K=2$ critical Boolean networks can be
understood as resulting from the network being composed of the order
of $N^{1/3}$ relevant nodes arranged in $\sim \ln N$ components, with
the number of components of size $l$ being Poisson distributed with a
mean $1/l$. Only the proportion $N^{-1/3}$ of all nodes is not frozen,
and those nonfrozen nodes that are not relevant sit in trees rooted in
relevant components.

The calculation indicates that there are several quantities that show
a scaling with $N$. Among these are the number of nonfrozen nodes, the
number of relevant nodes, and the number of nonfrozen nodes with
two nonfrozen inputs. Evaluating these scaling properties to more
detail could be the next step in understanding Kauffman networks.

I thank F. Greil and T. Mihaljev for comments on the manuscript.

\bibliography{../fgDT_RbnBib.bib, ../asynchron/ArbnBib.bib,../k1/PRL_k1.bib}

\end{document}